# BEYOND SOCIAL MEDIA ANALOGUES

Gregory M. Dickinson*

*The steady flow of social media cases to the Supreme Court reveals a nation reworking its fundamental relationship with technology. The cases raise a host of questions ranging from difficult to impossible: how to nurture a vibrant public square when a few tech giants dominate the flow of information, how social media can be at the same time free from conformist groupthink and protected against harmful disinformation campaigns, and how government and industry can cooperate on such problems without devolving toward censorship.*

*To such profound questions, this Essay offers a comparatively modest contribution—what not to do. Always the lawyer's instinct is toward analogy, considering what has come before and how it reveals what should come next. Almost invariably, that is the right choice. The law's cautious evolution protects society from disruptive change. But almost is not always, and with social media, disruptive change is already upon us. Using social media laws from Texas and Florida as a case study, this Essay suggests that social media's distinct features render it poorly suited to analysis by analogy and argues that courts should instead shift their attention toward crafting legal doctrines targeted to address social media's unique ills.*

## INTRODUCTION

It is nearly impossible to overstate the importance of the shift in the 1990s from the traditional author-publisher-distributor model of publication to the decentralized, internet-based modes of publication common today. The old gatekeepers—newspapers, book publishers, and television and radio broadcasters who chose what was worthy of publication—have lost their power. Now, anyone with an internet connection can share her thoughts with all who are willing to listen.[1] Americans have a love-hate relationship with

---

* Copyright © 2024 by Gregory M. Dickinson, Assistant Professor of Law and, by courtesy, Computer Science, St. Thomas University; Nonresidential Fellow, Stanford Law School, Program in Law, Science & Technology; J.D., Harvard Law School. Thanks to Rebeca Martinez for her tireless contributions and excellent research assistance.

1 *See generally* Kate Klonick, *The New Governors: The People, Rules, and Processes Governing Online Speech*, 131 HARV. L. REV. 1598, 1599–613 (2018) (tracing the rise of online platforms to become the new the free-flowing social media speech that followed the shift. Democratized publication has opened the public forum to a greater number of voices than ever before, but, at the same time, given rise to a host of difficult and deeply divisive questions about the proper role of government and online platforms in public debate.

Gone are the days when a few trusted arbiters of speech controlled the news that Americans consumed.[2] The old-guard institutions are still there, of course, but they do not shape public discourse with the same authority. The *New York Times* might report a fresh story, but it will never have the last word, and perhaps not even the most important one. Its voice is supplemented by innumerable others, often to the public's benefit, as online commentators explore new angles and late-breaking nuances. But these additional voices are decentralized, unverified, often uninformed, and sometimes even deceptive and malign.[3] They have the capacity to harm as well as to help. As a result, free-speech scholars are embroiled in a set of crucial debates about how the shift to online platforms will affect public discourse and democratic government: what, if anything, can be done to preserve a virtual public square free from both threats of targeted misinformation and government censorship;[4] how to protect Americans

---

gatekeepers of speech); Eugene Volokh, *Cheap Speech and What It Will Do*, 104 YALE L.J. 1805, 1806–07 (1995) (noting that historically the right to free speech has favored popular or well-funded ideas, but predicting that new information technologies would dramatically reduce the costs of distributing speech and create a more diverse and democratic environment).

　　[2]　*See generally* MARTHA MINOW, SAVING THE NEWS: WHY THE CONSTITUTION CALLS FOR GOVERNMENT ACTION TO PRESERVE FREEDOM OF SPEECH 10–21 (2021) (describing the decline of traditional journalism in the United States and attributing that decline in part to the rise of social media, which disrupted newspaper revenue streams); *see also* Gregory M. Dickinson, *Journalism in the Age of Clickbait*, 66 HOWARD L.J. 191 (2022) (discussing Minow's analysis and suggesting consideration also of the role that users' media-consumption preferences play in undermining traditional news publications).

　　[3]　*See, e.g.*, David E. Sanger & Steven Lee Myers, *China Uses A.I. to Spread Lies About U.S. Fire*, N.Y. TIMES, Sept. 11, 2023, at A1 (describing China's use of AI to create fake images of Hawaii fires for propaganda in the U.S.); Steven Lee Myers & Sheera Frenkel, *Exploding Online, Disinformation Is Now a Fixture of U.S. Politics*, N.Y. TIMES, Oct. 21, 2022, at A1 (recounting various political disinformation campaigns); Julian E. Barnes, *The Next Target in Putin's War: U.S. Support for Ukraine, Officials Say*, N.Y. TIMES, Oct. 3, 2023, at A6 (describing Russian spy agencies' use of social media to support U.S. political candidates opposed to continuing support for Ukraine).

　　[4]　*See, e.g.*, Erin C. Carroll, *A Free Press Without Democracy*, 56 U.C. DAVIS L. REV. 289, 303 (2022) (discussing the threat that autocratic regimes pose to press freedom); Julie E. Cohen, *Tailoring Election Regulation: The Platform Is the Frame*, 4 GEO. L. TECH. REV. 641, 655 (2020) (evaluating constitutionality of legislation aimed at election misinformation); Joseph Blocher, *Free Speech and Justified True Belief*, 133 HARV. L. REV. 439, 443–44 (2019) (urging a framework of First Amendment protection centered around justified true belief as a way to counter danger to free speech posed by overaggressive response to misinformation); Daniel Susser, Beate Roessler & Helen Nissenbaum, *Online Manipulation: Hidden Influences in a Digital World*, 4 GEO. L. TECH. REV. 1, 27 (2019) (noting the power of digital technologies to manipulate users, but also the

against cyberbullying, identity theft, revenge porn, and data-privacy intrusions;[5] and how to restore trust in public and private institutions despite a polarized political climate rife with claims and counterclaims of "fake news" and with governmental efforts to control controverted messages.[6] Scholars will wrestle with these questions for years to come as they percolate through the nation's court systems.

This Essay steps back from those debates to make a more fundamental point: Be cautious with analogy. Whether discussing online information, data privacy, platform governance, online governance, or the host of other questions raised by new publication technologies, lawyers' first move is to consider what has come before. How did law respond to the printing press? To the telegraph? To radio, television, or MP3? Legal doctrines and lines of precedent can be borrowed from earlier analogues to govern new circumstances without the disruption and expense of crafting a new legal solution from whole cloth every time a new technology emerges. But care must be taken to confirm the analogy fits, lest resort to history does more harm than good.

This Essay highlights the difficulty of analogy in the social media context by considering a pair of laws in Texas and Florida that aim to protect Americans' freedom of expression by requiring social media platforms to host user speech without regard to its viewpoint. Part I presents an overview of the statutes at issue, the legal challenges they have faced in the Fifth and Eleventh Circuits, and their pathway to the Supreme Court. Part II discusses the Supreme Court's compelled-speech precedents, which have guided

---

difficulty of regulating such technology given the imprecise boundary between persuasion and coercion).

[5] *See, e.g.*, Danielle Keats Citron, *A New Compact for Sexual Privacy*, 62 WM. & MARY L. REV. 1763, 1766–73 (2021) (describing the constant corporate surveillance of our digital lives and areas where current privacy law is not suited to address such harms); Danielle Keats Citron, *Sexual Privacy*, 128 YALE L.J. 1870 (2019) (discussing how networked technologies have facilitated various forms of sexual privacy violations); Margaret B. Kwoka, *FOIA, Inc.*, 65 DUKE L.J. 1361, 1376–401 (2016) (describing the industry of "information resellers" who request data under the Freedom of Information Act (FOIA) to resell for profit).

[6] *See* Aziz Huq & Tom Ginsburg, *How to Lose a Constitutional Democracy*, 65 UCLA L. REV. 78, 130–36 (2018) (discussing the process by which liberal democracies can regress, which includes the erosion of independent journalism and media); RonNell Andersen Jones & Sonja R. West, *The U.S. Supreme Court's Characterizations of the Press: An Empirical Study*, 100 N.C. L. REV. 375, 390–407 (2022) (documenting a trend toward less favorable discussions of the press in Supreme Court decisions); Mary Anne Franks, *Beyond the Public Square: Imagining Digital Democracy*, 131 YALE L.J.F. 427, 428 (2021) (urging a more considered and intentional structuring of the virtual world, as "the public square has historically tended to reinforce legal and social hierarchies of race, gender, class, and ability rather than foster radically democratic and inclusive dialogue"); Thomas E. Kadri, *Digital Gatekeepers*, 99 TEX. L. REV. 951, 951–57 (2021) (suggesting governmental action to limit the power of private, "gatekeep[ing]" entities to control the flow of internet data); Eugene Volokh, *Anti-Libel Injunctions*, 168 U. PA. L. REV. 73, 73–80 (2019) (describing how traditional libel remedies might apply to online contexts).

appellate court reasoning by way of analogy to organizing parades, leafletting at shopping malls, and recruiting on law-school campuses, and criticizes precedent-based reasoning as poorly suited to multifunction online platforms, whose features defy analogy to any single category of compelled-speech cases. Finally, Part III concedes the difficulty of legal analysis from first principles, without the benefit of analogy, yet reluctantly suggests that courts must be open to crafting and employing new doctrines to evaluate social media restrictions given the stark differences between online platforms and their physical-world predecessors.

I

THE SUPREME QUESTION

Social media are aflame with accusations and counteraccusations: Liberals have a strangle-hold on American news! Conservatives are driving a fascist insurrection! Silicon Valley elites silence the voice of the people! Red-staters are book-burning simpletons! For good and for ill, ideas on social media spread like wildfire—especially emotionally charged, half-true ones. The power of online speech puts speech itself at the center of public debate, with rhetoric at a fever pitch.

Concern is mounting on the left that our health, well-being, and even democracy itself may be in danger if something is not done to control the spread of misleading and hateful online speech.[7] In the wake of the COVID-19 pandemic and 2020 election, Washington State considered a legal measure that would have criminalized false, election-related speech;[8] and both California[9] and New York[10] have recently adopted laws that require social media companies to submit periodic reports to state officials regarding their strategies for policing online hate speech, misinformation, and other

---

[7] *See* Steven Lee Myers & Sheera Frenkel, *Exploding Online, Disinformation Is Now a Fixture of U.S. Politics*, N.Y. TIMES, Oct. 21, 2022, at A1 (discussing the spread of disinformation online regarding COVID-19, the 2020 U.S. presidential election, and the Russo-Ukrainian war, and highlighting difficulties in managing its impact on society); *see also* Tiffany Hsu, *Covid Misinformation Snowballs, Exasperating Doctors*, N.Y. TIMES, Jan. 2, 2023, at B1 (examining social media companies' approaches to limiting the spread of false information online); Elizabeth Dwoskin, *Misinformation on Facebook Got Six Times More Clicks than Factual News During the 2020 Election, Study Says*, WASH. POST, Sept. 5, 2021, at A7 (discussing a study by researchers at New York University that found greater engagement with posts containing misinformation supporting far-left and far-right political positions); *see also* Russell L. Weaver, *Remedies for "Disinformation"*, 55 U. PAC. L. REV. 185 (2024) (identifying the impacts of disinformation on the internet and proposing possible solutions to target its spread).

[8] S.B. 5843, 67th Leg., Reg. Sess. (Wash. 2022) (proposing the criminalization of "knowingly mak[ing] false statements or claims regarding the election process [and] election results" if made for the purpose of undermining the election and "directed to inciting . . . imminent lawless action").

[9] A.B. 587, 2021–2022 Leg., Reg. Sess. (Cal. 2022).

[10] 7865-A, 2021–2022 Leg., Reg. Sess. (N.Y. 2021).

enumerated categories of problematic content.[11]

On the right, the concern is the opposite—that efforts to control the spread of information online will prevent ordinary Americans from communicating their ideas. For the most strident, the issue is one of principle: Free citizens—even misguided, confused, or hateful ones—possess an innate right to speak their minds.[12] For others,[13] it is a pragmatic calculation: Free expression is a critical barrier against governmental overreach and concentration of power. Thus driven both by principle and pragmatics, Florida[14] and Texas[15] have enacted new laws to protect free expression by treating social media companies as common carriers and requiring them to host all users' speech regardless of viewpoint.

Were the stakes not so very high, we might all stop for a moment and smile: Flag-burning,[16] free-speech-loving Democrats of yesteryear[17]

---

[11] The New York law faces legal challenge by critics who argue that it includes viewpoint-based restrictions and unlawfully compels speech. The U.S. District Court for the Southern District of New York agreed and granted online platforms challenging the law a preliminary injunction barring enforcement, reasoning that the law "places Plaintiffs in the incongruous position of stating that they promote an explicit 'pro-free speech' ethos, but also requires them to enact a policy allowing users to complain about 'hateful conduct' as defined by the state." Volokh v. James, No. 22 CV 10195, 2023 WL 1991435, at *7 (S.D.N.Y. Feb. 14, 2023). The California law has also faced opposition on First Amendment grounds, with critics equating the law to a requirement that "the New York Times . . . explain which stories it publishes." Cat Zakrzewski, *New California Law Likely to Set off Fight Over Social Media Moderation*, WASH. POST (Sept. 13, 2022) (quoting Adam Kovacevich, the C.E.O. of an industry coalition that includes Meta and Google), https://www.washingtonpost.com/technology/2022/09/13/california-social-network-transparency [https://perma.cc/7VUR-2AHB].

[12] *See, e.g.*, Alan K. Chen, *Free Speech, Rational Deliberation, and Some Truth About Lies*, 62 WM. & MARY L. REV. 357, 358 (2020) (arguing that free speech doctrine should limit the state's "ability to control the way we emotionally experience ideas, beliefs, and even facts . . . ," even if those facts could be categorized as being "fake news").

[13] *See, e.g.*, Kathleen McGarvey Hidy, *Social Media Use and Viewpoint Discrimination: A First Amendment Judicial Tightrope Walk with Rights and Risks Hanging in the Balance*, 102 MARQ. L. REV. 1045, 1060 (2019) ("[T]he Free Speech Clause helps produce informed opinions among members of the public, who are then able to influence the choices of a government that, through words and deeds, will reflect its electoral mandate."); *see also* Russell L. Weaver, *Should Congress (or, for that Matter, a New Federal Authority) Regulate Social Media?*, 58 U. GA. L. REV. (forthcoming 2024) (recognizing the drawbacks of unregulated disinformation, but also emphasizing the difficulties in implementing information regulation in the United States).

[14] S.B. 7072, 2021 Leg. (Fla. 2021) (reasoning that social media platforms have become the new public square and barring social media companies from "deplatforming" political candidates, censoring or "shadow banning" certain election-related speech, and imposing disclosure obligations on platforms to explain their standards for content moderation).

[15] H.B. 20, 87th Leg., 2d Sess. (Tex. 2021) (requiring social media companies to disclose their content-moderation practices and treating them as common carriers, prohibiting them from engaging in viewpoint-based censorship of their users' speech).

[16] *See* Texas v. Johnson, 491 U.S. 397 (1989) (finding that a defendant's burning of an American flag constituted an expressive act protected by the First Amendment); *see also* United States v. Eichman, 496 U.S. 310 (1990) (same).

[17] *See Berkeley Leftists Burn Flags, in Miniature, to Mark Decision*, L.A. TIMES, June 23,

condemn today's charged online rhetoric as a danger to American democracy, while law-and-order Republicans[18] charge into the breach to defend free-speech ramparts built by the sweat and blood of 1970s radicals. But this is no time to revel in the irony. The great online speech debate is upon us—flush with substance, compelling arguments all around, and real consequences for individual freedom of expression and American democratic government.

Now the debate has landed in the Supreme Court, which has agreed to consider challenges to the Florida[19] and Texas[20] common-carrier laws. The laws are a response to long-standing concerns[21] that the mainstream media are biased and to recent high-profile instances of online censorship, such as Facebook's blocking discussion of the theory that COVID-19 escaped from

---

1989, at 23 (reporting on a communist youth brigade at Berkeley burning American flags to celebrate the Supreme Court decision regarding a First Amendment right to do so); Tom Kenworthy, *Flag Amendment Sent to House Floor*, WASH. POST, June 20, 1990, at A14 (describing opposition by leaders of the Democratic party to the constitutional ban on flag burning, motivated by a concern to protect free speech); David E. Rosenbaum, *In Recurring Debate, House Votes to Ban Flag-Burning*, N.Y. TIMES, June 25, 1999, at A18 (same).

[18] *See* Flag Protection Act of 1989, 18 U.S.C. § 700 (Republican-sponsored act that criminalized "mutilat[ing], defac[ing], physically defil[ing], burn[ing], maintain[ing] on the floor or ground, or trampl[ing] upon any flag of the United States"). The Supreme Court declared the Act unconstitutional in *United States v. Eichman*, 496 U.S. 310 (1990); s*ee also* Ruth Marcus, *Nude Dancing Covered by the Constitution?*, WASH. POST, Jan. 6, 1991, at A1 (reporting sentiment in Indiana, prior to the Supreme Court's decision in *Barnes v. Glen Theatre, Inc.*, 501 U.S. 560 (1991), that it "does not compute" that "[y]ou can dance nude but you can't pray in school," and that "'[t]hose guys who wrote the Constitution would be twirling' at the thought that conduct like nude dancing would be viewed as free speech").

[19] NetChoice, LLC v. Att'y Gen., Fla., 34 F.4th 1196 (11th Cir. 2022), *cert. granted in part sub nom.* Moody v. NetChoice, LLC, 216 L. Ed. 2d 1313 (Sept. 29, 2023), *and cert. denied* NetChoice, LLC v. Moody, 144 S. Ct. 69 (2023).

[20] NetChoice, LLC v. Paxton, 49 F.4th 439 (5th Cir. 2022), *cert. granted in part* NetChoice, LLC v. Paxton, 216 L. Ed. 2d 1313 (Sept. 29, 2023).

[21] *See, e.g.*, Nikolas Lanum, *Twitter, Facebook, Google Have Repeatedly Censored Conservatives Despite Liberal Doubts*, FOX NEWS (Mar. 29, 2022), https://www.foxnews.com/media/twitter-facebook-google-censored-conservatives-big-tech-suspension [https://perma.cc/7TG2-SGB6] ("A pattern has emerged of right-leaning voices being censored far more often than those on the left."); *see also* Brian Flood, *Big Tech's 'Secondhand Censorship' Shields Conservatives from Information at Alarming Rate, Study Shows*, FOX NEWS (Jul. 21, 2022), https://www.foxnews.com/media/big-techs-secondhand-censorship-shields-conservatives-information-alarming-rate-study-shows [https://perma.cc/TJ9U-BLJG] (discussing the findings of a study conducted by MRC Free Speech America, a conservative organization that claims to track effects of secondhand censorship). The MRC study finds that Facebook, Twitter, YouTube, Instagram, TikTok, LinkedIn, and Spotify kept an "astonishing" amount of information from users in 2022. The authors argue that "America is increasingly outraged by the manner in which radical Big Tech leftists are censoring conservative and Christian leaders . . . on nearly every major social media platform." Brian Bradley & Gabriela Pariseau, *The Secondhand Censorship Effect: The Real Impact of Big Tech's Thought-Policing*, MRC FREE SPEECH AMERICA (Jul. 20, 2022),
https://censortrack.org/secondhand-censorship-effect-real-impact-big-techs-thought-policing [https://perma.cc/UNL9-HPU5].

a virology laboratory in Wuhan, China,[22] and Twitter's disabling of the New York Post's account after it published a report on the contents of Hunter Biden's laptop.[23]

The laws are being challenged by NetChoice, a tech industry trade association whose members include Twitter (now rebranded as X), Meta, Pinterest, and Google.[24] The platforms' public response has been to argue that laws targeting their content-moderation practices are unnecessary because their practices are not biased. Testifying before Congress last year, the CEOs of Facebook, Twitter, and Google argued that their content-moderation policies are entirely neutral, aimed at harmful and offensive content regardless of its political valence.[25] But their legal challenges to Texas's and Florida's social media laws take a slightly different tack. There, they argue that even were their content-moderation decisions in part politically motivated, their decisions about what content to promote and what content to censor are protected from interference by their own First Amendment interests in free expression.[26]

---

[22]    *See* Steven W. Mosher, *Don't Buy China's Story: The Coronavirus May Have Leaked from a Lab*, N.Y. POST (Feb. 22, 2020), https://nypost.com/2020/02/22/dont-buy-chinas-story-the-coronavirus-may-have-leaked-from-a-lab [https://perma.cc/3E6Y-MZZW] (highlighting facts supporting speculation that the virus could have originated in a laboratory in Wuhan, China); *see also* Newley Purnell, *Facebook Ends Ban on Posts Asserting Covid-19 Was Man-Made*, WALL ST. J. (May 27, 2021), https://www.wsj.com/articles/facebook-ends-ban-on-posts-asserting-covid-19-was-man-made-11622094890 [https://perma.cc/3DFL-3Y4V] (reporting on Facebook's policy shift concerning a ban on any posts that asserted the coronavirus was manufactured or created in a laboratory).

[23]    *See* Farnoush Amiri & Barbara Ortutay, *Ex-Twitter Execs Deny Pressure to Block Hunter Biden Story*, ASSOCIATED PRESS (Feb. 8, 2023), https://apnews.com/article/technology-politics-united-states-government-us-republican-party-business-6e34ad121a1e52892b782b0b7c0e59c3 [https://perma.cc/4WZ8-A38U] (reporting that Twitter executives admitted mistake in blocking the New York Post's story regarding the contents of Hunter Biden's laptop, which broke during the build-up to the 2020 presidential election, citing Twitter's blocking the content as an attempt to "avoid repeating the mistakes of 2016").

[24]    *See generally* NetChoice, *About Us*, NETCHOICE.ORG, https://netchoice.org/about/#our-mission [https://perma.cc/AV2U-TX8U] (describing its mission as "to make the Internet safe for free enterprise and free expression").

[25]    "Suppressing content on the basis of political viewpoint or preventing people from seeing what matters most to them is directly contrary to Facebook's mission and our business objectives." *Facebook, Social Media Privacy, and the Use and Abuse of Data: Joint Hearing Before the Comm. on Com., Sci., and Transp. & the Comm. on the Judiciary*, 115th Cong. 2 (2018) (statement of Mark Zuckerberg, CEO of Meta Platforms, Inc.). Google's CEO, Sundar Pichai, also stated within his written testimony, "We approach our work without political bias, full stop. To do otherwise would be contrary to both our business interests and our mission." *Does Section 230's Sweeping Immunity Enable Big Tech Bad Behavior?*, *Testimony for the Record for the Comm. on Com., Sci., and Transp.*, 116th Cong. (2020) (statement of Sundar Pichai, CEO of Alphabet, Inc.).

[26]    *See* NetChoice, LLC v. Paxton, 49 F.4th 439, 463 (5th Cir. 2022) ("[The Platforms] argue that Section 7 interferes with their speech by infringing their 'right to exercise editorial discretion' . . . [and that] 'editorial discretion' is a separate, freestanding category of First-Amendment-protected expression."); NetChoice, LLC v. Att'y Gen., Fla., 34 F.4th 1196, 1203 (11th Cir. 2022)

That argument is the crux of the debate. Americans demand freedom to express their views online, even misinformed or disfavored ones. Yet platforms have their own interests to defend. They are trying to build online environments that people actually want to visit. Conspiracy theorists and agitators shouting, spewing nonsense, and sowing virtual chaos tend to drive away business and attract the attention of lawmakers. The problem is that online platforms are private entities, with an obvious interest in curating user content, but they are so large and so dominant that they have also become potential chokepoints of public discourse.

The consequences of the debate could not be more serious, but the cases are also a First Amendment scholar's dream come true. They set the stage for a fascinating battle over the First Amendment's text, spirit, and precedent. Recall the text of the First Amendment, which provides that "*Congress* shall make no law . . . abridging the freedom of speech."[27] It says nothing about private social media companies. What happens when private companies, rather than the government, are the ones that censor user speech? May the legislature intervene to further the First Amendment's goal of free discourse among citizens? And what about social media companies' own interest in expression?[28] No platform wants to be known as the internet's cesspool. They have brands and reputations to uphold.

## II
### ANALOGY, PRECEDENT, AND PITFALLS

Under the Court's compelled-association cases, the viability of the Florida and Texas laws depends on whether one thinks Twitter is more like a parade or a shopping mall—the circumstances of two leading First Amendment decisions from the Supreme Court. First, in *Hurley v. Irish-American Gay, Lesbian & Bisexual Group of Boston*,[29] the Court forbade Massachusetts from forcing private parade organizers to include a gay, lesbian, and bisexual-pride group in their parade. The Court acknowledged the general validity of Massachusetts's public accommodations law, which prohibits discrimination on the basis of sexual orientation. But it went on to

---

(holding that social media companies were "private actors" and finding it likely that their "so-called 'content-moderation' decisions constitute protected exercises of editorial judgment, and that the provisions of the new Florida law that restrict large platforms' ability to engage in content moderation unconstitutionally burden that prerogative." (quoting Manhattan Comm. Access Corp. v. Halleck, 139 S.Ct 1921, 1926 (2019)).

27   U.S. CONST. amend. I (emphasis added).

28   It may initially sound strange to call platforms' censoring or downgrading certain bits of content "expression," but the expressive character of editorial discretion is well established in Supreme Court precedent. *See* Miami Herald Publ'g Co. v. Tornillo, 418 U.S. 241, 258 (1974) (observing that a newspaper's decisions about what content to publish constitute editorial control and judgment protected by the First Amendment).

29   515 U.S. 557 (1995).

conclude that because "every participating unit [in a parade] affects the message conveyed by the private organizers," enforcement of the law to require the parade organizers to include the group would "alter the expressive content of their parade" and therefore violate "the fundamental rule of protection under the First Amendment, that a speaker has the autonomy to choose the content of his own message."[30]

In reaching that result, *Hurley* distinguished *Turner Broadcasting System, Inc. v. FCC*, decided only a year earlier, which upheld a requirement that cable companies devote some of their channels to local broadcast stations.[31] The *Hurley* Court reasoned that parades do not consist of "individual, unrelated segments"—rather, they convey a "common theme," making it difficult for a parade organizer to disavow association with one of the parade units.[32] The Court contrasted parades with cable television programming, where it is unlikely that "cable viewers would assume that the broadcast stations carried on a cable system convey ideas or messages endorsed by the cable operator" itself.[33] Another point distinguished the cases too. *Turner* had been justified in part by the concern that some speakers, namely local broadcasters, would be "destroyed" and unable to convey their messages if the market-dominant cable operators were not required to carry them.[34] No such concern was present in *Hurley*, where the parade organizer did not "enjoy the capacity to 'silence the voice of competing speakers,'" for any other group could simply ask the city for a parade permit of its own.[35]

Another guidepost is *PruneYard Shopping Center v. Robins*, in which the Supreme Court considered a provision of California's state constitution that creates a right for members of the public to engage in "free speech and petition rights" in privately-owned shopping centers.[36] The appellant in that case was a shopping mall owner who enforced a general policy against noncommercial expressive activity to exclude from his property a group of students that was distributing pamphlets.[37] The owner argued that California's constitutional provision was inconsistent with his right under the First Amendment to "not to be forced by the State to use his property as a forum for the speech of others."[38] Writing for the Court, Justice Rehnquist rejected the owner's arguments, noting that the property was "not limited to the [owner's] personal use . . . ," but was "instead a business establishment that is open to the public . . . ," and that, given the vast size of the shopping center and its numerous separate shops, "the views expressed by members of the public in passing out pamphlets . . . [would] not likely be identified with those of the owner."[39]

---

[30] *Id.* at 572–73; *see* MASS. GEN. LAWS, ch. 272, § 98 (2016).
[31] 512 U.S. 622 (1994).
[32] Hurley v. Irish-Am. Gay, Lesbian, & Bisexual Grp. of Bos., Inc., 515 U.S. 557, 576 (1995).
[33] *Id.* (quoting Turner Broad. Sys., Inc. v. FCC, 512 U.S. 622, 655 (1994)).
[34] *Id.* at 577; *Turner*, 512 U.S. at 662.
[35] *Hurley*, 515 U.S. at 578 (quoting *Turner*, 512 U.S. at 656).
[36] 447 U.S. 74, 76 (1980).
[37] *Id.* at 77.
[38] *Id.* at 74.

So which is Twitter, a parade or a shopping mall? The Eleventh Circuit[40] thought a parade and the Fifth Circuit[41] a shopping mall. But nobody really knows. Online platforms are different from both and more complicated than either. Like parade organizers, online platforms tailor their content feeds to attract eyeballs and create atmospheres that are consistent with their brands. They have reputations to uphold and user bases to build. But almost all of their moderation decisions are made algorithmically, to maximize views and clicks—not manually to express a message. And, like shopping malls, online platforms are businesses open to the public; readers will attribute posts to their third-party speakers, not the platforms' owners. Additionally, unlike a moving parade where "disclaimers would be quite curious,"[42] platform disclaimers of third-party views are not only possible, but already in place.[43]

The dictates of America's precedent-based legal system tell us that this is the question to ask:[44] Which past legal decisions most closely mirror the facts of the present case, and how did the courts decide those cases? That approach makes sense. Courts aim to decide like cases alike, and what better way to do so than by adhering to past decisions? The law must, of course, evolve as litigants bring novel disputes before the court, but many formerly open issues have by now been resolved, and those decisions hold the force of law as binding precedent. Through this approach, abstract philosophizing about "the Good" is left to legislatures, future questions are left for future courts, and a present court need only resolve the dispute before it, typically by applying established case law.[45] That is the bread and butter of precedent-based legal systems and it is exactly what the Eleventh Circuit chose to do, what the Fifth Circuit should have done, and what the Supreme Court will be sorely tempted to do. But it should not.

---

39   *Id.* at 87.
40   *See* NetChoice, LLC v. Att'y Gen., Fla., 34 F.4th 1196, 1213 (11th Cir. 2022) ("Just as the parade organizer exercises editorial judgment when it refuses to include in its lineup groups with whose messages it disagrees . . . [p]latforms employ editorial judgment to convey some messages but not others and thereby cultivate different types of communities that appeal to different groups.").
41   *See* NetChoice, LLC v. Paxton, 49 F.4th 439, 464 (5th Cir. 2022) (observing that "[p]latforms cannot invoke 'editorial discretion' as if uttering some sort of First Amendment talisman to protect their censorship" and "[w]ere it otherwise, the shopping mall in *PruneYard* . . . could have changed the outcomes of [the] case[] by simply asserting a desire to exercise 'editorial discretion'").
42   *Hurley*, 515 U.S. at 576.
43   *See Terms of Service*, X (Sept. 29, 2023), https://twitter.com/en/tos [https://perma.cc/P8GN-T8X9] ("We do not endorse, support, represent or guarantee the completeness, truthfulness, accuracy, or reliability of any Content or communications posted via the Services or endorse any opinions expressed via the Services."); *see also Terms of Service*, FACEBOOK (July 26, 2022), https://www.facebook.com/terms [https://perma.cc/BL4R-GWAP] ("We do not control or direct what people and others do or say, and we are not responsible for their actions or conduct (whether online or offline) or any content they share (including offensive, inappropriate, obscene, unlawful, and other objectionable content)."); *Terms of Service*, YOUTUBE (Jan. 5, 2022), https://www.youtube.com/static?template=terms [https://perma.cc/LN28-QABD] ("Content is the responsibility of the person or entity that provides it to the Service. YouTube is under no obligation to host or serve Content.").
44   *See, e.g.*, WILLIAM BLACKSTONE, COMMENTARIES ON THE LAWS OF ENGLAND 69 (1771) ("For it is an established rule to abide by former precedents, where the same points come again in litigation: as well to keep the scale of justice even and steady, and not liable to waver with every new judge's opinion . . . ."); EDWARDO COKE, THE FOURTH PART OF THE INSTITUTES OF THE LAWS OF ENGLAND; CONCERNING THE JURISDICTION OF COURTS 109 (1817) ("[L]et us now peruse our ancient authors, for out of the old fields must come the new corn.").

The easy way out for the Supreme Court would be to resolve the NetChoice cases narrowly, by pure application of precedent: The Florida and Texas laws violate the First Amendment because they require social media companies to alter the expressive content on their platforms. Or, alternatively, they are consistent with the First Amendment because social media companies have opened their platforms to the public and the views of platform users will not likely be attributed to platform owners.[46] This approach exemplifies classic judicial minimalism.[47] Stay within the contours of the present case and avoid speculating about the future because you might very well be wrong.

However, taking the easy path to resolve these cases could be a mistake for a few reasons. First, a basic application of precedent either to approve or disapprove of the Florida and Texas social media laws would leave in place the deep tension in the Court's compelled-speech cases. To take one leading example, in *Hurley*, the Court prohibited Massachusetts from forcing parade organizers to include a gay, lesbian, and bisexual group in their midst,[48] whereas in *Rumsfeld v. Forum for Academic and Institutional Rights*,[49] the

---

[45] *See* THE FEDERALIST NO. 78, at 496 (Alexander Hamilton) (Benjamin F. Wright ed., Harvard Univ. Press 1974) ("To avoid an arbitrary discretion in the courts, it is indispensable that they should be bound down by strict rules and precedents, which serve to define and point out their duty in every particular case that comes before them . . . .").

[46] *Cf., e.g.*, 303 Creative LLC v. Elenis, 143 S. Ct. 2298, 2305–06 (2023) (discussing the important role of public accommodations laws in combating discrimination, particularly when an enterprise exercises "something like monopoly power," but also noting that no law is "immune from the demands of the Constitution" and that "public accommodations statutes can sweep too broadly when deployed to compel speech").

[47] *See generally* Cass R. Sunstein, *Burkean Minimalism*, 105 MICH. L. REV. 353, 356 (2006) (exploring the concept of Burkean minimalism, rooted in the philosophy of Edmund Burke, which involves "respecting settled judicial doctrine, but also deferring to traditions").

[48] Hurley v. Irish-Am. Gay, Lesbian, & Bisexual Grp. of Bos., Inc., 515 U.S. 557 (1995).

[49] 547 U.S. 47, 69–70 (2006) (holding that military recruiters are outside members, not affiliated with the law school, "who come onto campus for the limited purpose of trying to hire

Court upheld a federal law that forced law schools to allow military recruiters onto their campuses. In short, the Court's cases appear simultaneously to guarantee First Amendment protection against government-compelled speech, but also recognize the government's power under *PruneYard* and related cases to force private property owners to host third-party speech—even speech they disagree with, and even where speech is a central part of their own mission. Various theoretical resolutions of the tension have been proposed,[50] but none has been clearly embraced by the Court.

Sure, a decision that Twitter is more like a parade than a shopping mall, or vice versa, would bring an end to the dispute. But it would be just one more entry in the free- or compelled-association ledger. Little progress would be made toward reconciling the two branches of the Court's cases. What the public really needs to know is why. What legal principle justifies government compulsion of private parties to host speech they disapprove of and what are the limits of that principle? Merely to pick a side, as did the Eleventh Circuit, by analogizing platforms to shopping malls and parades,[51] would perpetuate the Court's failure to provide a coherent theoretical account of its decisions.

Second, a terse decision under contested precedents might appear to be driven by the Court's political priors rather than established law—an ironic result given that a goal of Robertsian judicial minimalism is to produce

---

students—not to become members of the school's expressive association," and, therefore, a law school has no First Amendment-protected editorial discretion to deny military recruiters the same access given to other recruiters at the school).

50  *See, e.g.*, Nicholas Bramble, *Ill Telecommunications: How Internet Infrastructure Providers Lose First Amendment Protection*, 17 MICH. TELECOMM. & TECH. L. REV. 67, 81–83 (2010) (contrasting *Rumsfeld* and *Hurley* and attempting to reconcile their holdings by looking to whether a requirement "sufficiently interfere[s]" with a speaker's message or causes others' messages to be reasonably imputed to them); Richard A. Epstein, *Takings, Exclusivity and Speech: The Legacy of PruneYard v. Robins*, 64 U. CHI. L. REV. 21, 22–28, 53–56 (1997) (urging a takings law "derivative of or dependent on private law conceptions of property" such that government-authorized physical occupation of private property would constitute a per se taking and not require recourse to any balancing test); James M. Gottry, *Just Shoot Me: Public Accommodation Anti-Discrimination Laws Take Aim at First Amendment Freedom of Speech*, 64 VAND. L. REV. 961, 988 (2011) (attempting to reconcile tension between *Hurley* and *Rumsfeld* on the ground that the law schools in *Rumsfeld* "had a choice to continue their educational mission without interference by simply forfeiting federal funding—a solution that several schools implemented"); Ashutosh Bhagwat, *Why Social Media Platforms Are Not Common Carriers*, 2 J. FREE SPEECH L. 127, 149–50 (2022) (noting the tension between cases like *Rumsfeld* and *Hurley* and suggesting the key difference is whether conduct is expressive or nonexpressive).

51  NetChoice, LLC v. Att'y Gen., Fla., 34 F.4th 1196, 1213, 1215 (11th Cir. 2022) (likening social media platforms to parades in that "social-media companies are in the business of delivering curated compilations of speech created . . . by others," while distinguishing social media platforms from a *PruneYard*-type shopping mall in that the owner of the shopping center, unlike social media platforms, never asserted that "access to [the shopping center in order to protest] might affect the shopping center owner's exercise of his own right to speak . . . ." (quoting Pac. Gas & Elec. Co. v. Pub. Utils. Comm'n of Cal., Inc., 475 U.S. 1, 12 (1986))).

apolitical, balls-and-strikes judging as often as possible.[52] But judicial umpiring can only work if we know what game we are playing. The last four decades have seen the Court call compelled-association claims as balls and strikes without ever explaining the rules of the game.[53] Resolving another case without a theory of how to distinguish permissible from impermissible speech compulsion will continue that trend into the politically salient context of online speech. The result could be a public backlash à la *Kelo*[54] or *Roe*[55] that risks undermining the Court's credibility.[56]

Third, platforms are not monolithic. Part of what makes platforms so attractive is their wide-ranging functionality.[57] Much like blogs, platforms

---

[52] *See Confirmation Hearing on the Nomination of John G. Roberts, Jr. To Be Chief Justice of the United States Before the S. Comm. on the Judiciary*, 109th Cong. 55–56 (2005) (statement of John G. Roberts, Jr.) ("The role of an umpire and a judge is critical. They make sure everybody plays by the rules . . . and I will remember that it's my job to call balls and strikes, and not to pitch or bat."); *see also* Dobbs v. Jackson Women's Health Org., 142 S. Ct. 2228, 2311 (2022) (Roberts, C.J., concurring) ("If it is not necessary to decide more to dispose of a case, then it is necessary *not* to decide more.").

[53] Many have attempted to reconcile these judicial decisions, but in that process, draw different key distinctions between the two cases. Some suggest the focus should be on the expressive nature of the platform itself, as the parade in *Hurley* is inherently expressive and the law school recruitment in *Rumsfeld* and the shopping mall in *PruneYard* are not. Bhagwat, *supra* note 50, at 149–50. Others suggest a "reasonable imputation" test that asserts the key determination to be whether the accommodation of speech on a host's platform sufficiently interferes with the host's own message to where the participant's message will be reasonably imputed to that of the host. Bramble, *supra* note 50, at 83. Some scholars have provided other supporting distinctions, such as the host's ability to choose, specifically, on the choice to speak on the matter or avoid the participant's inclusion entirely. This was the case in *Rumsfeld*, as the law school could simply forfeit federal funding in order to deny the military recruitment, whereas *Hurley* centered around the choice to not put forth a specific viewpoint, a choice that would be denied to the parade if they were forced to include the participants. Gottry, *supra* note 50, at 988.

[54] Kelo v. City of New London, 545 U.S. 469, 483–84 (2005) (holding that the eminent domain taking of a woman's home qualified as a "public use" because the city's economic development plan predicted benefits to the community overall if the woman's property were transferred to a private developer).

[55] Roe v. Wade, 410 U.S. 113 (1973) (interpreting the federal Constitution to create a right to abortion services).

[56] *See* T. R. Reid, *Missouri Condemnation No Longer So Imminent*, WASH. POST (Sept. 6, 2005), https://www.washingtonpost.com/archive/politics/2005/09/06/missouri-condemnation-no-longer-so-imminent/3569d07b-91b7-491d-926e-7058eb5b0b87 [https://perma.cc/CR2R-XYLR] (describing the backlash that occurred after the *Kelo* decision, including multiple bipartisan legislation proposals aimed at limiting the kind of seizure the Court's decision justified); Michael Klarman, *Marriage Equality and Political Backlash*, N.Y. TIMES (July 1, 2015, 6:05 PM), https://www.nytimes.com/roomfordebate/2013/03/26/civil-rights-decisions-in-courts-or-legislatures/marriage-equality-and-political-backlash [https://perma.cc/KSU6-PKCH] ("*Roe*'s aggressive defense of abortion rights fostered a right-to-life movement that fundamentally reshaped American politics and arguably made abortion reform more contentious and resistant to compromise.").

[57] *See* Eugene Volokh, *Treating Social Media Platforms Like Common Carriers?*, 1 J. FREE SPEECH L. 377, 408–09 (2021) (considering multiple distinct functions that social media platforms provide and suggesting that some functions, including its "recommendation function," by which

allow users to author content and post it publicly to their individual pages. These individual user pages also serve another function. Members of the public who visit a user's page can post their own comments or pictures in response, thereby expanding and livening (or, just as likely, disrupting and frustrating) the user's original message. To help users and commenters craft eyeball-attracting content, platforms pitch in by providing GIF databases and photo-editing functionality. And, because making money is important too, platforms curate and repackage the body of material posted by their users into real-time content feeds that are customized for each user and prominently displayed (mixed with targeted ads) on the main page of the app or website.

In short, social media platforms are like the blind men and the elephant.[58] How you perceive them depends where you look. If you grab Twitter by its content-feed trunk, it may appear something like a newspaper, magazine, or parade, whose private, expressive content the Court has been keen to protect. It selects (albeit algorithmically) specific content for display in its content feeds, leaving the rest of its vast content stores on the cutting-room floor. But if instead you run your hand along Twitter's broad, wall-like sides, that is, the personal pages of its users, you will find something very like a public square or shopping mall, where the Court has permitted more invasive governmental action. Wherever you examine the social media elephant you will find different functionality requiring its own distinct First Amendment analysis.

Social media's cornucopia of functionality makes it poorly suited to minimalist resolution as parade- or shopping-mall-like under existing precedent. Flagging user posts as inflammatory is different from removing them, is different from issuing warnings to posters, is different from shadow banning them so that only they can see their own content, and is different from deplatforming them entirely.[59] The analysis of each could change depending on whether the user's posts are moderated on the user's own page, her comment section, the platform's curated content feed, or all three.

---

the platform emphasizes selected posts, are less amenable to common-carrier treatment than others, especially their "hosting function," which allows people to post content to the platform, similar to phone companies and the postal service).

[58] An ancient Hindu parable about a group of blind men who touch a different part of an elephant and disagree as to what it is—an incomplete representation of a whole. 1 ENCYCLOPEDIA OF PERCEPTION 492 (E. Bruce Goldstein ed., 2010).

[59] *See* Eric Goldman, *Content Moderation Remedies*, 28 MICH. TECH. L. REV. 1, 23–24, 28 (2021) (developing a taxonomy of options, including shadow banning and deplatforming, available to platforms to reduce undesired content); Tarleton Gillespie, *Reduction / Borderline Content / Shadowbanning*, 24 YALE J.L. & TECH. 476, 487 (2022) (defining shadow banning to include any of various actions surreptitiously taken by a platform to make a user's content less visible, while "leaving the user thinking they are participating normally").

## III

## CHARTING A DIFFERENT COURSE

In principle, the Court could just split up a platform's functionality into pieces and classify each separately—deciding this bit of functionality is parade-like, that one shopping-mall-like, and so on. But that approach would only make more glaring the lack of a unifying theory to explain *why* parades and shopping malls are treated differently in the first place. Why does the First Amendment permit the government to compel a private shopping mall owner or private law school to serve as a platform for unwanted speech, but not a private parade organizer?

It is not hard to imagine rationales.[60] Perhaps the concern is that in the context of parades (but not shopping malls), the public will misperceive the situation and interpret parade organizers to endorse the messages of the speakers to whom they are forced to open their platforms.[61] Or perhaps parades are "speaking organizations" entitled to greater freedom from compulsion than less speech-centric entities.[62] Legal scholars have offered more than a few suggestions, and at times the Supreme Court has hinted at the importance of one or the other.[63] But no unifying rationale is fully consistent with the case law and none has been expressly endorsed by the

---

[60] *See* Gottry, *supra* note 50, at 988 (reconciling *Hurley* and *Rumsfeld* by emphasizing the host's "choice" in speaking or lack thereof, as *Hurley* centered around the protection of the parade-speaker's choice to not put forth a specific viewpoint, while *Rumsfeld* did not remove the choice from the law school to speak on that case's pertinent issue, as "schools had a choice to continue their educational mission without interference by simply forfeiting federal funding—a solution that several schools implemented"); Joel Timmer, *Promoting and Infringing Free Speech? Net Neutrality and the First Amendment*, 71 FED. COMMC'N L.J. 1, 14–16 (2018) (reconciling the decisions in *Rumsfeld* and *Hurley* by focusing on the likelihood that the message at issue could be identified with the host's message, where a parade inherently conveys a message, but a law school allowing the military to recruit students conveys no message, nor does it compel or deny the school's speech).

[61] *See* Hurley v. Irish-Am. Gay, Lesbian & Bisexual Grp. of Bos., 515 U.S. 557, 572–73 (1995) (observing that parades are symbolic expressions meant to convey certain messages and that "every participating unit affects the message conveyed by the private organizers").

[62] *See* Volokh, *supra* note 57, at 423–28 (pointing out that every participating unit affects the coherent speech product that a parade attempts to convey, and that allowing participants that the parade organizers disavowed would affect the parade's own message, a key characteristic missing in other First Amendment cases such as *PruneYard*, *Turner*, and *Rumsfeld*); Bhagwat, *supra* note 50, at 144 (highlighting the Supreme Court's emphasis on the expressive nature of parades in deciding in *Hurley* that a law requiring parade organizers to include viewpoints they disagreed with would violate their right against compelled speech).

[63] *See, e.g.*, Glickman v. Wileman Bros. & Elliott, 521 U.S. 457, 469–70 (1997) (finding that the requirement that fruit growers contribute funds to a joint advertising fund did not constitute compelled speech); Rumsfeld v. F. for Acad. & Institutional Rts., Inc., 547 U.S. 47, 66 (2006) (holding that a mandate to permit recruiters on campus was permissible because it did not regulate inherently expressive conduct); 303 Creative LLC v. Elenis, 143 S. Ct. 2298, 2320 n.6, 2321–22 (2023) (allowing that "public accommodations laws may sometimes touch on speech incidentally," but concluding that requiring a website designer to create a wedding website for a gay couple would impermissibly compel speech in violation of the First Amendment).

Court. Now is the time for the Court to change that. The Court should end its journey down the easy path—the unsystematized application of precedent to not-truly-monolithic platform functionality—and instead develop a rationale that will explain its decision as to each type of speaker and each bit of functionality.

There are many paths the Court could take, but one of the most intriguing and widely discussed[64] is the common-carrier proposal—the idea that social media platforms, like telecommunications companies, public utilities, transportation companies and the like, might be so critical to public well-being that they must provide service equally to all comers, despite companies' typical freedom to accept or reject customers as they see fit.[65]

---

64  *See* Sarah S. Seo, Note, *Failed Analogies: Justice Thomas's Concurrence in* Biden v. Knight First Amendment Institute, 32 FORDHAM INTELL. PROP. MEDIA & ENT. L.J. 1070, 1098–99 (2022) (arguing that Justice Thomas's rationales for equating social media platforms to common carriers are flawed and that common-carrier treatment would frustrate current precedent and impair the free market for computer services); Bobby Allyn, *Justice Clarence Thomas Takes Aim at Tech and Its Power 'To Cut Off Speech'*, NPR (Apr. 5, 2021, 1:27 PM), https://www.npr.org/2021/04/05/984440891/justice-clarence-thomas-takes-aims-at-tech-and-its-power-to-cut-off-speech [https://perma.cc/3X9D-JHP6] (characterizing common-carrier treatment of social media platforms as "a fringe view popular among partisan conservatives but not shared by federal regulatory agencies and the Supreme Court's own precedents"); Josh Gerstein, *Justice Thomas Grumbles over Trump's Social Media Ban*, POLITICO (Apr. 5, 2021, 10:58 AM), https://www.politico.com/news/2021/04/05/justice-clarence-thomas-trump-twitter-ban-479046 [https://perma.cc/Y89S-PLC8] (analogizing a common-carrier requirement for social media to requirements that force business owners "to accept customers regardless of race or religion"); George F. Will, Opinion, *Clarence Thomas Is Right. Big Tech Will Have Its Day in Court, Eventually*, WASH. POST (Apr. 9, 2021, 8:00 AM), https://www.washingtonpost.com/opinions/clarence-thomas-is-right-the-supreme-court-should-referee-big-tech/2021/04/08/99459f12-9885-11eb-a6d0-13d207aadb78_story.html [https://perma.cc/TF5U-9FGP] (suggesting that Justice Thomas's concurrence promoting a common-carrier analysis is "perhaps too certain" that the marketplace dominance of social media companies such as Google and Facebook presents a sufficient barrier "to entry [to] 'entrench' such companies against competitors"). *But see* Eugene Volokh, *Justice Thomas Suggests Rethinking Legal Status of Digital Platforms*, REASON: VOLOKH CONSPIRACY (Apr. 5, 2021, 10:49 AM), https://reason.com/volokh/2021/04/05/justice-thomas-suggests-rethinking-of-legal-status-of-digital-platforms [https://perma.cc/Q64R-QCEX] (suggesting Justice Thomas's concurrence was "not arguing that platforms are already generally common carriers or government actors under existing legal principles," but anticipating potential legislative approaches that would treat social media platforms as common carriers).

65  The intuition underlying common-carrier doctrine is simple enough: Companies can become so powerful or so important that they should be required to accept all willing customers. But the details are contested, and some scholars question whether any discrete, coherent concept can be extracted from the cases. *See* Blake E. Reid, *Uncommon Carriage*, 76 STAN. L. REV. 89, 107–08 (2024) ("[D]espite the apparent . . . appetite for theories of common carriage, there is no consensus as to the contours of common carriage law—or whether common carriage represents a discrete concept."). Interested readers should consult, among others sources, James B. Speta, *A Common Carrier Approach to Internet Interconnection*, 54 FED. COMMC'N L.J. 225 (2002) (suggesting that lessons from common-carrier regulation should be placed on the internet); Daniel T. Deacon, *Common Carrier Essentialism and the Emerging Common Law of Internet Regulation*, 67 ADMIN.

In a 2021 decision, *Biden v. Knight First Amendment Institute at Columbia University*, Justice Clarence Thomas raised the possibility that a similar principle should govern social media.[66] The Knight Institute challenged former President Trump's authority to block members of the public from responding to his tweets. Although the Supreme Court dismissed the challenge as moot given Trump's election loss, Justice Thomas took the opportunity to comment on the legal challenges surrounding online speech and social media. He observed that social media platforms "provide avenues for historically unprecedented amounts of speech," but that never before has "control of so much speech [been] in the hands of a few private parties."[67] He went on to suggest that perhaps social media companies, like railroads and telephone companies, should be required to accept all users equally, without regard to their viewpoints.[68] Most people, after all, would not want telephone companies to deny service to individuals based on their political or religious views. Perhaps online platforms should be treated similarly.

That is the approach taken by the Florida and Texas social media statutes, which include legislative findings that online platforms are common carriers and thus subject to special duties to the public to convey their users' messages without regard to viewpoint.[69] Since Justice Thomas suggested the approach and the Florida and Texas legislatures adopted it, common-carrier doctrine would be a natural fit for the Court to employ in evaluating those states' laws.[70]

But common-carrier doctrine is a difficult path. The Court's compelled-association cases may lack a unifying theoretical principle, but the line of precedent is well developed and comparatively robust. By contrast, the common-carrier doctrine is rarely discussed in the Court's decisions and

---

L. REV. 133 (2015) (assessing the FCC approach to internet regulation and claiming that common-carrier essentialism is not stable enough to develop policy); Adam Candeub, *Bargaining for Free Speech: Common Carriage, Network Neutrality, and Section 230*, 22 YALE J.L. & TECH. 391 (2020) (interpreting common carriage and network regulation through the lens of good exchange); Volokh, *supra* note 57; Christopher S. Yoo, *The First Amendment, Common Carriers, and Public Accommodations: Net Neutrality, Digital Platforms, and Privacy*, 1 J. FREE SPEECH L. 463 (2021) (analyzing what counts as a common carrier and what are the First Amendment implications).

66   141 S. Ct. 1220, 1222–23 (2021) (Thomas, J., concurring).
67   *Id.*
68   *Id.* at 1225.
69   *See* S.B. 7072, 2021 Leg., Reg. Sess. (Fla. 2021) ("Social media platforms hold a unique place in preserving first amendment protections for all Floridians and should be treated similarly to common carriers."); H.B. 20, 87th Leg., Reg. Sess. (Tex. 2021) ("[S]ocial media platforms function as common carriers, are affected with a public interest, are central public forums for public debate . . . and [that] social media platforms with the largest number of users are common carriers by virtue of their market dominance.").
70   *Cf.* Epstein, *supra* note 50, at 50–53 (observing that an entity should not ordinarily be subject to enhanced limitations on its property rights merely because of its size, but excepting instances of monopoly power, where "one group exerts a dominant power over some given resource").

dramatically undertheorized.[71] Developing a full-fledged theory of common-carrier doctrine as related to First Amendment freedom of association and compelled-speech cases would be an enormous undertaking.

Moreover, common-carrier treatment would be a costly solution to the social media problem. Treating entities as common carriers interferes with their freedom to respond to market demands and reduces consumer welfare, undermines corporate profitability and spending on innovation by increasing compliance and litigation costs, and imposes licensing requirements and other barriers that solidify market leaders and benefit incumbents.[72] Common-carriage requirements are accepted as necessary for the public good but are burdensome and applied sparingly, typically where industry incumbents hold significant market power and some economic or social barrier makes a market-driven solution unlikely.

Yet common-carrier doctrine may be a better mode of analysis. Unlike the shopping mall–parade distinction, the common-carrier doctrine goes to the very heart of the problem: Have large social media platforms become so dominant and so critical to modern public discourse that they can be forced to convey messages expressing viewpoints they disagree with, or have they not? That is the real issue, and common-carrier doctrine is an analytical tool suited to address it. No one cares whether online platforms are parades or shopping malls or purple petunias. What people want to know is whether platforms' role in societal discourse is so pivotal and their market power so strong as to warrant their being forced to carry the public's messages.

CONCLUSION

Whether the government may compel private parties to host speech is a question that has bedeviled commentators for decades. The stars have aligned for the Court to take up the question this term. And the challenges to the Florida and Texas laws are only two of a multitude of social media cases on the horizon. The Court has already agreed next term to review a pair of decisions from the Sixth[73] and Ninth Circuits[74] that consider whether government officials may block users from commenting on their Twitter feeds. Even more recently, a Louisiana trial court issued, the Fifth Circuit modified, and the Supreme Court thereafter stayed a preliminary injunction barring Biden administration officials from communicating with social media platforms to coordinate the identification and removal of purported misinformation.[75] The rush of social media questions shows no sign of slowing. Whichever way the cases go on the merits, judges should resist the temptation to fall back on precedential distinctions and timeworn analogies. Online platforms are a generational technology that defies analogy and requires fresh consideration via appropriate doctrinal tools.

---

[71] *See* Yoo, *supra* note 65 (discussing the legal history of common carriers and public accommodation laws and the scope of their respective First Amendment rights); *see also* Speta, *supra* note 65 (examining the similarities between traditional common carriers and the internet); Deacon, *supra* note 65 (asserting that the approach to common-carrier law is unstable in its application to emerging internet industries, relying on case-by-case factual adjudications); Reid, *supra* note 65 (analyzing the lack of coherence in common-carrier legal theory, as well as commenting on the First Amendment implications if internet platforms are equated to common carriers).

[72] *See* Biden v. Knight First Amend. Inst. at Columbia Univ., 141 S. Ct. 1220, 1224 (2021) (Thomas, J., concurring) (mentioning the astronomical profit margins of social media platforms, and how there is a substantial barrier for competitors in entering this high-stakes market); Daniel F. Spulber & Christopher S. Yoo, *Mandating Access to Telecom and the Internet: The Hidden Side of* Trinko, 107 COLUM. L. REV. 1822, 1899 (2007) (emphasizing the importance of competition amongst networks, as standardization "decreases welfare by reducing product variety" and that "incompatible networks may simply represent the natural outgrowth of heterogenous consumer preferences"); Susan P. Crawford, *The Radio and the Internet*, 23 BERKELEY TECH. L.J. 933, 953–56 (2008) (highlighting the economic benefits and overall importance of preserving competition amongst internet servicers).

---

73  *See* Lindke v. Freed, 37 F.4th 1199 (6th Cir. 2022), *cert. granted*, 143 S. Ct. 1780 (2023) (No. 22-611).

74  *See* O'Conner-Ratcliff v. Garnier, 41 F.4th 1158 (9th Cir. 2022), *cert. granted*, 143 S. Ct. 1779 (2023) (No. 22-324).

75  Missouri v. Biden, No. 3:22-CV-01213, 2023 WL 4335270 (W.D. La. July 4, 2023), *aff'd in part, rev'd in part*, 83 F.4th 350 (5th Cir. 2023), *cert. granted sub nom.* Murthy v. Missouri, No. 23-411, 2023 WL 6935337 (U.S. Oct. 20, 2023); *see also* Steven L. Myers & David McCabe, *Federal Judge Limits Biden Officials' Contacts with Social Media Sites*, N.Y. TIMES (July 4, 2023), https://www.nytimes.com/2023/07/04/business/federal-judge-biden-social-media.html [https://perma.cc/4NP6-Y872] (explaining the district judge's grant of a preliminary injunction and the First Amendment implications on the limits of online speech).